\documentclass[a4paper,UKenglish,cleveref, autoref, thm-restate]{lipics-v2021}
\nolinenumbers
\usepackage{amsthm}
\usepackage{mathrsfs}
\usepackage{graphicx} 
\usepackage{algorithm}
\usepackage{algpseudocode}
\usepackage{url}
\title{A contraction-recursive algorithm for treewidth}

\author{Hisao Tamaki}
{Meiji University, Kawasaki Japan} {hisao.tamaki@gmail.com} 
{https://orcid.org/0000-0001-7566-8505}{}

\authorrunning{Hisao Tamaki}
\Copyright{Hisao Tamaki} 

\ccsdesc[500]{Theory of computation~Graph algorithms analysis}

\keywords{graph algorithm, treewidth, exact computation, 
BT dynamic programming, contraction, certifying algorithms}

\category{} 

\relatedversion{} 

\EventEditors{John Q. Open and Joan R. Access}
\EventNoEds{2}
\EventLongTitle{42nd Conference on Very Important Topics (CVIT 2016)}
\EventShortTitle{CVIT 2016}
\EventAcronym{CVIT}
\EventYear{2016}
\EventDate{December 24--27, 2016}
\EventLocation{Little Whinging, United Kingdom}
\EventLogo{}
\SeriesVolume{42}
\ArticleNo{23}

\newcommand{\defic}{\mathop{defic}}

\newcommand{\improve}{\mathop{improve}}
\newcommand{\expand}{\mathop{expand}}

\newcommand{\contractPMCs}{\mathop{contractPMCs}}
\newcommand{\uncPMCs}{\mathop{uncontractPMCs}}

\newcommand{\tw}{\mathop{tw}}
\newcommand{\calT}{{\mathscr{T}}}

\newcommand{\calA}{{\cal A}}
\newcommand{\calC}{{\cal C}}
\newcommand{\calK}{{\cal K}}
\newcommand{\calS}{{\cal S}}
\newcommand{\calB}{{\cal B}}
\newcommand{\calX}{{\cal X}}
\newcommand{\calF}{{\cal F}}
\newcommand{\calP}{{\cal P}}
\newcommand{\return}{\textbf{return }}

\newcommand{\fillg}{\mathop{fill}}

\newcommand{\minimalize}{\mathop{minimalize}}
\newcommand{\minopt}{\mathop{minimalize\_optimally}}

\newcommand{\finish}{\mathop{finish}}
\newcommand{\importPMCs}{\mathop{importPMCs}}
\newcommand{\usefulPMCs}{\mathop{usefulPMCs}}
\newcommand{\width}{\mathop{width}}
\newcommand{\budget}{\mathop{budget}}
\newcommand{\searchNewFeasible}{\mathop{searchNewFeasible}}

\begin{document}
\maketitle
\begin{abstract}
Let $\tw(G)$ denote the treewidth of graph $G$.
Given a graph $G$ and a positive integer $k$ such that $\tw(G) \leq k + 1$, 
we are to decide if $\tw(G) \leq k$. 
We give a certifying algorithm RTW ("R" for recursive) for this task:
it returns one or more tree-decompositions of $G$ of width $\leq k$ if the answer is YES
and a minimal contraction $H$ of $G$ such that $\tw(H) > k$ otherwise.
Starting from a greedy upper bound on $\tw(G)$ and repeatedly improving the
upper bound by this algorithm, we obtain $\tw(G)$ with certificates.

RTW uses a heuristic variant of
Tamaki's PID algorithm for treewidth (ESA2017), which we call
HPID. Informally speaking, PID builds potential subtrees of tree-decompositions of 
width $\leq k$ in a bottom up manner, until such a tree-decomposition is
constructed or the set of potential subtrees is exhausted without success.
HPID uses the same method of generating a new subtree from existing ones but
with a different generation order which is not intended for exhaustion but 
for quick generation 
of a full tree-decomposition when possible.
RTW, given $G$ and $k$, interleaves the execution of HPID with
recursive calls on $G /e$ for edges $e$ of $G$,
where $G / e$ denotes the graph obtained from $G$ by contracting
edge $e$. If we find that $\tw(G / e) > k$, then we have
$\tw(G) > k$ with the same certificate. If we find that $\tw(G / e) \leq k$,
we "uncontract" the bags of the certifying tree-decompositions of $G / e$ 
into bags of $G$ and feed them to HPID to help progress.
If the question is not resolved after the recursive calls are made for
all edges, we finish HPID in an exhaustive mode.
If it turns out that $\tw(G) >  k$,  
then $G$ is a certificate for $\tw(G') > k$ for every $G'$ of which $G$ is a contraction, 
because we have found $\tw(G / e) \leq k$ for every edge $e$ of $G$. 
This final round of HPID guarantees the correctness of the algorithm, 
while its practical
efficiency derives from our methods of "uncontracting" bags of
tree-decompositions of $G / e$ to useful bags of $G$, as well as of
exploiting those bags in HPID.

Experiments show that our algorithm drastically extends the scope of practically solvable instances. In particular, when applied to the 100 instances in the PACE 2017 bonus set, the number of instances solved by our implementation on a typical laptop, with the timeout of 100, 1000, and 10000 seconds per instance, are 72, 92, and 98 respectively, while
these numbers are 11, 38, and 68 for Tamaki's PID solver and
65, 82, and 85 for his new solver (SEA 2022).
\end{abstract}
\maketitle

\section{Introduction}
Treewidth is a
graph parameter introduced and extensively studied in the graph minor theory \cite{robertson1986graph}. 
A \emph{tree-decomposition} of graph $G$ is a tree
with each node labeled by a vertex set of $G$,
called a {\em bag}, satisfying certain conditions
(see Section~\ref{sec:prelim}) so that those bags form a tree-structured system of vertex-separators of $G$. The width $w(T)$ of a tree-decomposition $T$
is the maximum cardinality of a bag in $T$ minus one and the treewidth $\tw(G)$ of graph $G$ is the smallest $k$ such that there is a tree-decomposition of $G$ of width $k$.

The impact of the notion of treewidth on the design of combinatorial algorithms is profound: there are a huge number of NP-hard graph problems that are
known to be tractable when parameterized by treewidth: they admit
an algorithm with running time $f(k)n^{O(1)}$, where $n$ is the number of vertices, $k$ is the treewidth of the given graph, and $f$ is some typically exponential function (see \cite{cygan2015parameterized}, for example). 
Those algorithms typically perform dynamic programming based on the system of separators
provided by the tree-decomposition. To make such algorithms practically useful, we need to compute the treewidth, or a good approximation of the treewidth, together with an associated tree-decomposition.

Computing the treewidth $\tw(G)$ of a given graph $G$ is NP-complete \cite{arnborg1987complexity}, but is fixed-parameter tractable \cite{robertson1986graph, bodlaender1996linear}. In particular, the algorithm due to Bodlaender \cite{bodlaender1996linear} runs in time linear in the graph size with a factor of $2^{O(\tw(G)^3)}$.
Unfortunately, this algorithm does not seem to run efficiently in practice.

In more practical approaches to treewidth computation, triangulations of graphs
play an important role. A \emph{triangulation} of graph $G$ is a chordal graph
$H$ with $V(G) = V(H)$ and $E(G) \subseteq E(H)$. For every tree-decomposition $T$ of $G$,
filling every bag of $T$ into a clique gives a triangulation of $G$. Conversely,
for every triangulation $H$ of $G$, there is a tree-decomposition of $G$ in which
every bag is a maximal clique of $H$.
Through this characterization of tree-decompositions in terms of triangulations, we can
enumerate all relevant tree-decompositions by going through the total orderings on the vertex set, as each total ordering defines a triangulation for which the ordering is a perfect elimination order (see \cite{bodlaender2010treewidth}, for example). 
Practical algorithms in the early stage of treewidth research performed a branch-and-bound search over these total orderings \cite{bodlaender2010treewidth}. 
Dynamic programming on this search space results in a $2^n n^{O(1)}$ time algorithms \cite{bodlaender2006exact}, which works well in practice for graphs with a small number of vertices.
It should also be noted that classical upper bound algorithms, such as min-deg or min-fill, which heuristically choose a single vertex ordering defining a tree-decomposition, are fast and often give a good approximation of the treewidth in a practical sense \cite{bodlaender2010treewidth}.

Another important link between chordal graphs and
treewidth computation was established by Bouchitt\'{e} and Todinca \cite{bouchitte2001treewidth}. They introduced the notion of potential maximal cliques (PMCs,
see below in "Our approach" paragraph for a definition)
and gave an efficient dynamic programming algorithm working on PMCs (BT dynamic programming) to
find a minimal triangulation of the given graph that corresponds to an optimal tree-decomposition.
They showed that their algorithm runs in polynomial time for many special classes of graphs. BT dynamic programming is also used in an exponential time algorithm for treewidth that runs in time
$O(1.7549^{n})$ \cite{fomin2012treewidth}.

BT dynamic programming had been considered mostly of theoretical interest until 2017, when Tamaki presented its positive-instance driven (PID) variant, which runs fast in practice and significantly outperforms previously implemented treewidth algorithms \cite{tamaki2019positive}. Further efforts
on treewidth computation based on or around his approach have been made since then, 
with some incremental successes \cite{tamaki2019heuristic,tamaki2019computing,tamaki2021heuristic, althaus2021ontamaki}.

In his most recent work \cite{tamaki:LIPIcs.SEA.2022.17}, Tamaki introduced another approach to treewidth computation, based on the use of contractions to compute tight lower bounds on the treewidth. For edge $e$ of graph $G$,
the \emph{contraction} of $G$ by $e$, denoted by $G / e$,
is a graph obtained from $G$ by replacing $e$ by a new single vertex $v_e$ and
let $v_e$ be adjacent to all neighbors of the ends of $e$ in $V(G) \setminus e$. 
A graph $H$ is a \emph{contraction} of $G$ if $H$ is obtained from $G$
by zero ore more successive contractions by edges.
It is well-known and easy to see that
$\tw(H) \leq \tw(G)$ for every contraction $H$ of $G$. This fact has been used
to quickly compute reasonably good lower bounds on the treewidth of a graph, typically to be used in branch-and-bound algorithms mentioned above \cite{bodlaender2010treewidth,bodlaender2011treewidth}. 
Tamaki \cite{tamaki:LIPIcs.SEA.2022.17} gave a
heuristic method of successively improving contraction based lower bounds which, 
together with a separate heuristic method for upper bounds, quite often succeeds in
computing the exact treewidth of instances that are hard to solve for previously published solvers.

\subsubsection*{Our approach}
Our approach is based on the observation that contractions are useful not only for
computing lower bounds but also for computing upper bounds.
Suppose we have a tree-decomposition $T$ of $G / e$ of width $k$ 
for some edge $e = \{u, v\}$ of $k$. Let $v_e$ be the vertex to which $e$ contracts.
Replacing each bag $X$ of $T$ by $X'$, where $X' = X \setminus \{v_e\} \cup \{u, v\}$ if
$v_e \in X$ and $X' = X$ otherwise, we obtain a tree-decomposition $T'$ of
$G$ of width $\leq k + 1$, which we call the \emph{uncontraction} of $T$. 
In a fortunate case where every bag $X$ of $T$
with $v_e \in X$ has $|X| \leq k$, the width of $T'$ is $k$.
To increase the chance of having such fortunate cases, we deal with
a set of tree-decompositions rather than a single tree-decomposition.
We represent such a set of tree-decompositions by a set of potential maximal cliques as follows.

A vertex set of $G$ is a \emph{potential maximal clique} (PMC for short) if
it is a maximal clique of some minimal triangulation of $G$. 
Let $\Pi(G)$ denote the set of all PMCs of $G$. 
For each $\Pi \subseteq \Pi(G)$, let $\calT_{\Pi}(G)$ denote the
set of all tree-decompositions of $G$ whose bags all belong to $\Pi$.
Let $\tw_{\Pi}(G)$ denote the smallest $k$ such that there is a tree-decomposition
in $\calT_{\Pi}(G)$ of width $k$; we set $\tw_{\Pi}(G) = \infty$ if
$\calT_{\Pi}(G) = \emptyset$.
Bouchitt\'{e} and Todinca \cite{bouchitte2001treewidth}
showed that $\calT_{\Pi(G)}(G)$ contains a tree-decomposition of width $\tw(G)$ and developed a dynamic programming algorithm (BT dynamic programming) to find such a tree-decomposition. 
Indeed, as Tamaki \cite{tamaki2019heuristic} noted, BT dynamic programming can be used for arbitrary $\Pi \subseteq \Pi(G)$ to compute $\tw_{\Pi}(G)$ in time linear in $|\Pi|$ and polynomial in $|V(G)|$. 

A set of PMCs is a particularly effective representation of a set of tree-decompositions
for our purposes, because BT dynamic programming can be used to work on
$\Pi \subseteq \Pi(G)$ and find a tree-decomposition in $\calT_\Pi(G)$
that minimizes a variety of width measures based on bag weights. 
In our situation, suppose we have
$\Pi \subseteq \Pi(G / e)$ such that $\tw_\Pi(G / e) = k$.
Using appropriate bag weights, we can use BT dynamic programming to decide if
$\calT_\Pi(G / e)$ contains $T$ such that the uncontraction $T'$
of $T$ has width $k$ and find one if it exists. 

These observation suggests a recursive algorithm for improving an upper bound on treewidth.
Given graph $G$ and $k$ such that $\tw(G) \leq k + 1$, the task is to
decide if $\tw(G) \leq k$. Our algorithm certifies the YES answer by
$\Pi \subseteq \Pi(G)$ with $\tw_\Pi(G) \leq k$. It uses heuristic methods to find
such $\Pi$ and, when this goal is hard to achieve, recursively solve
the question if $\tw(G / e) \leq k$ for edge $e$ of $G$.
Unless $\tw(G / e) = k + 1$ and hence $\tw(G) = k + 1$,
the recursive call returns $\Pi \subseteq \Pi(G / e)$ such that
$\tw_\Pi(G / e) \leq k$.  We use the method mentioned above to look for 
$T \in \calT_\Pi(G / r)$ whose uncontraction has width $\leq k$.
If we are successful, we are done for $G$. Even when this is not the case,
the uncontractions of tree-decompositions in $\calT_\Pi(G / e)$ may be useful
for our heuristic upper bound method in the following manner.

In \cite{tamaki2019heuristic}, Tamaki proposed a local search algorithm for treewidth
in which a solution is a set of PMCs rather than an individual tree-decomposition and
introduced several methods of expanding $\Pi \subseteq \Pi(G)$ into 
$\Pi' \supset \Pi$ in hope of having $\tw_{\Pi'}(G) < \tw_\Pi(G)$. His method compares favourably with existing heuristic algorithms but, like typical local search methods,
is prone to local optima.  To let the search escape from a local optimum, we would
like to inject "good" PMCs to the current set $\Pi$. It appears 
that tree-decompositions in $\calT_{\Pi'}(G / e)$ such that $\tw_{\Pi'} (G / e) \leq k$,
where $k = \tw_\Pi(G) - 1$,
are reasonable sources of such good PMCs: we uncontract $T \in \calT_{\Pi'}(G / e)$
into a tree-decomposition $T'$ of $G$ and extract PMCs of $G$ from $T'$.
Each such PMC appears in a tree-decomposition of width $\leq k + 1$ and
may appear in a tree-decomposition of width $\leq k$. It is also
important that $\Pi'$ is obtained, in a loose sense, independently of $\Pi$ and
not under the influence of the local optimum around which $\Pi$ stays.

Our algorithm for deciding if $\tw(G) \leq k$ interleaves the
execution of a local search algorithm with recursive calls on
$G / e$ for edges $e$ of $G$ and injects PMCs obtained from 
the results of the recursive calls. This process ends in 
either of the following three ways.
\begin{enumerate}
\item The local search succeeds in finding $\Pi$ with $\tw_\Pi(G) \leq k$.
\item A recursive call on $G / e$ finds that $\tw(G / e) = k + 1$:
we conclude that $\tw(G) = k + 1$ on the spot.
\item Recursive calls $G / e$ have been tried for all edges $e$ and 
it is still unknown if $\tw(G) \leq k$. We invoke a conventional
exact algorithm for treewidth to settle the question.
\end{enumerate}
Note that, when the algorithm concludes that $\tw(G) = k + 1$,
there must be a contraction $H$ of $G$ somewhere down in the recursion path from $G$
such that Case 3 applies and the exact computation shows
that $\tw(H) = k + 1$. In this case, $H$ is a minimal contraction
of $G$ that certifies $\tw(G) = k + 1$, as the recursive calls further down from
$H$ have shown $\tw(H / e) \leq k$ for every edge $e$ of $H$.

As the experiments in Section~\ref{sec:experiments} show, 
this approach drastically extends the scope of instances for which the exact treewidth can be computed in practice.

\paragraph*{Organization}
To quickly grasp the main ideas and contributions of this paper, it is suggested to read the following sections first: Section~\ref{sec:main} -- Main algorithm, Section~\ref{sec:uncontract} -- Uncontracting PMCs, Section~\ref{sec:contract} -- Contracting PMCs, and
Section~\ref{sec:experiments} -- Experiments  (about 9 pages in total including the
introduction), together with some parts of the preliminaries section as needed. 
Section~\ref{sec:hpid} describes some details of the local search algorithm we use, namely 
heuristic PID. Sections~\ref{sec:safe_sep},  \ref{sec:edge_order},
and \ref{sec:suppressed} describe additional techniques for speeding up the main algorithm. Sections~\ref{sec:conclusions} offers some concluding remarks.

The source code of the implementation of our algorithm used in the experiments is available at
\url{https://github.com/twalgor/RTW}.
\section{Preliminaries}
\label{sec:prelim}
\paragraph*{Graphs and treewidth}
In this paper, all graphs are simple, that is, without self loops or
parallel edges. Let $G$ be a graph.
We denote by $V(G)$ the vertex set
of $G$ and by $E(G)$ the edge set of $G$.
As $G$ is simple, each edge of $G$ is  
a subset of $V(G)$ with exactly two members that
are adjacent to each other in $G$.
The \emph{complete graph} on $V$, denoted
by $K(V)$, is a graph with vertex set $V$ 
in which every vertex is adjacent to all other vertices.
The subgraph of $G$ induced by $U \subseteq V(G)$ is denoted by 
$G[U]$.  We sometimes use an abbreviation $G \setminus U$ to
stand for $G[V(G) \setminus U]$. 
A vertex set $C \subseteq V(G)$ is a \emph{clique} of $G$ if
$G[C]$ is a complete graph.
For each $v \in V(G)$, $N_G(v)$ denotes the set of neighbors of $v$ in $G$:
$N_G(v) = \{u \in V(G) \mid \{u, v\} \in E(G)\}$.
For $U \subseteq V(G)$, the {\em open neighborhood of $U$ in $G$}, denoted
by $N_G(U)$,  is the set of vertices adjacent to some vertex in $U$ but not
belonging to $U$ itself: $N_G(U) = (\bigcup_{v \in U} N_G(v)) \setminus U$. 

We say that vertex set $C \subseteq V(G)$ is {\em connected in}
$G$ if, for every $u, v \in C$, there is a path in $G[C]$ between $u$
and $v$. It is a {\em connected component} or simply a {\em component} 
of $G$ if it is connected and is inclusion-wise maximal subject to this condition.
We denote by $\calC(G)$ the set of all components of $G$. 
When the graph $G$ is clear from the context, we denote $\calC(G[U])$ by
$\calC(U)$.
A vertex set $S \subseteq V(G)$ is a {\em separator} of $G$ if 
$G \setminus S$ has more than one component.
A graph is a \emph{cycle} if it is connected and every vertex
is adjacent to exactly two vertices. A graph is a \emph{forest}
if it does not have a cycle as a subgraph. A forest is
a \emph{tree} if it is connected.

A {\em tree-decomposition} of $G$ is a pair $(T, \calX)$ where $T$ is a tree
and $\calX$ is a family $\{X_i\}_{i \in V(T)}$ of vertex sets of $G$, indexed
by the nodes of $T$, such that the following 
three conditions are satisfied. We call 
each $X_i$ the {\em bag} at node $i$.   
\begin{enumerate}
  \item $\bigcup_{i \in V(T)} X_i = V(G)$.
  \item For each edge $\{u, v\} \in E(G)$, there is some $i \in V(T)$
  such that $u, v \in X_i$.
  \item For each $v \in V(G)$, the set of nodes $I_v = \{i \in V(T) \mid v \in
  X_i\} \subseteq V(T)$ is connected in $T$.
\end{enumerate}
The {\em width} of this tree-decomposition is $\max_{i \in V(T)} |X_i| - 1$.
The {\em treewidth} of $G$, denoted by $\tw(G)$ is the smallest $k$ such
that there is a tree-decomposition of $G$ of width $k$.

For each pair $(i, j)$ of adjacent
nodes of a tree-decomposition $(T, \calX)$ of $G$, let
$T(i, j)$ denote the subtree of $T$ consisting of nodes of $T$
reachable from $i$ without passing $j$ and let 
$V(i, j) = \bigcup_{k \in V(T(i, h))} X_k$.
Then, it is well-known and straightforward to show that
$X_i \cap X_j = V(i, j) \cap V(j, i)$ and there are no edges between
$V(i, j) \setminus V(j, i)$ and $V(j, i) \setminus V(i, j)$;
$X_i \cap X_j$ is a separator of $G$ unless $V(i, j) \subseteq V(j, i)$ or
$V(j, i) \subseteq V(j, i)$. We say that $T$ \emph{uses}
separator $S$ if there is an adjacent pair $(i, j)$ such that
$S = X_i \cap X_j$. In this paper, we assume $G$ is connected
whenever we consider a tree-decomposition of $G$.

In this paper, most tree-decompositions are such that $X_i = X_j$ only if $i = j$.
Because of this, we use a convention to view a tree-decomposition of $G$ as a tree $T$
whose nodes are bags (vertex sets) of $G$.

\paragraph*{Triangulations, minimal separators, and Potential maximal cliques}
Let $G$ be a graph and $S$ a separator of $G$. 
For distinct vertices $a, b \in V(G)$, 
$S$ is an {\em $a$-$b$
separator} if there is no path between $a$ and $b$ in $G \setminus S$; 
it is a 
{\em minimal $a$-$b$ separator} if it is an $a$-$b$ separator and
no proper subset of $S$ is an $a$-$b$ separator.
A separator is a {\em minimal separator} if it is a minimal $a$-$b$ separator
for some $a, b \in V(G)$.  

Graph $H$ is {\em chordal} if every induced cycle of $H$ has exactly three vertices. 
$H$ is a {\em triangulation of graph $G$} if it is chordal,
$V(G) = V(H)$, and
$E(G) \subseteq E(H)$. A triangulation $H$ of $G$ is {\em minimal}
if it there is no triangulation $H'$ of $G$ such that
$E(H')$ is a proper subset of $E(H)$.
It is known (see \cite{heggernes2006minimal} for example) that
if $H$ is a minimal triangulation of $G$ then every minimal separator of $H$
is a minimal separator of $G$. In fact, the set of minimal separators of $H$
is a maximal set of pairwise non-crossing minimal separators of $G$, where
two separators $S$ and $R$ \emph{cross each other} if at least two components
of $G \setminus S$ intersects $R$.

Triangulations and tree-decompositions are closely related.
For a tree-decomposition $T$ of $G$, let $\fillg(G, T)$ denote
the graph obtained from $G$ by filling every bag of $T$ into a clique.
Then, it is straightforward to see that $\fillg(G, T)$ is a triangulation of $G$.
Conversely, for each chordal graph $H$, consider a tree 
on the set $\calK$ of all maximal cliques of $H$ such that
if $X, Y \in \calK$ are adjacent to each other then
$X \cap Y$ is a minimal separator of $H$. Such a tree is called a 
\emph{clique tree} of $H$. It is straightforward to verify that
a clique tree $T$ of a triangulation $H$
of $G$ is a tree-decomposition of $G$ and that $\fillg(G, T) = H$.

We call a tree-decomposition $T$ of $G$ \emph{minimal} if it is a clique
tree of a minimal triangulation of $G$. It is clear that there is a minimal
tree-decomposition of $G$ of width $\tw(G)$, since for every tree-decomposition
$T$ of $G$, there is a minimal triangulation $H$ of $G$ that
is a subgraph of $\fillg(G, T)$ and every clique tree $T'$ of $H$
has $w(T') \leq w(T)$.

A vertex set $X \subseteq V(G)$ is a {\em potential maximal
clique}, PMC for short, of $G$, if $X$ is a maximal clique in 
some minimal triangulation of $G$.
We denote by $\Pi(G)$ the set of all potential maximal  
cliques of $G$. By definition, every bag of a minimal tree-decomposition
of $G$ belongs to $\Pi(G)$.

\paragraph*{Bouchitt\'{e}-Todinca dynamic programming}
For each $\Pi \subseteq \Pi(G)$, say that $\Pi$ \emph{admits} a tree-decomposition $T$
of $G$ if every bag of $T$ belongs to $\Pi$.
Let $\calT_\Pi(G)$ denote the
set of all tree-decompositions of $G$  that $\Pi$ admits and
let $\tw_\Pi(G)$ denote the smallest $k$ such that there is $T \in \calT_\Pi(G)$ of
width $k$; we set $\tw_\Pi(G) = \infty$ if $\calT_\Pi(G) = \emptyset$.
The treewidth algorithm of 
Bouchitt\'{e} and Todinca \cite{bouchitte2001treewidth} is based on the
observation that $\tw(G) = \tw_{\Pi(G)}(G)$. Given $G$, their algorithm first constructs $\Pi(G)$
and then search through $\calT_{\Pi(G)}(G)$ by dynamic programming (BT dynamic programming)
to find $T$ of width $\tw_{\Pi(G)}(G)$. As observed in \cite{tamaki2019heuristic},
BT dynamic programming can be used to compute $\tw_{\Pi}(G)$ for
an arbitrary subset $\Pi$ of $\Pi(G)$ to produce an upper bound on $\tw(G)$.
As we extensively use this idea, we describe how it works here.

Fix $\Pi \subseteq \Pi(G)$ such that $\calT_{\Pi}(G)$ is non-empty.
To formulate the recurrences in BT dynamic programming, we need
some definitions. A vertex set $B$ of $G$ is a \emph{block} if
$B$ is connected and either $N_G(B)$ is a minimal separator or is empty. 
As we are assuming that $G$ is connected, $B = V(G)$ in the latter case.
A \emph{partial tree-decomposition} of a block $B$ in $G$
is a tree-decomposition of $G[B \cup N_G(B)]$ that has a 
bag containing $N_G(B)$, called the \emph{root bag} of
this partial tree-decomposition. Note that a partial tree-decomposition
of block $V(G)$ is a tree-decomposition of $G$.
For graph $G$ and 
block $B$, let $\calP_{\Pi}(B, G)$ denote the set of
all partial tree-decompositions of $B$ in $G$ all of whose
bags belong to $\Pi$ and, when this set is non-empty,
let $\tw_{\Pi}(B, G)$ denote the smallest $k$ such that
there is $T \in \calP_{\Pi}(B, G)$ with $w(T) = k$;
if $\calP_{\Pi}(B, G)$ is empty we set $\tw_{\Pi}(B, G) = \infty$.

A PMC $X$ of $G$ is a \emph{cap} of block $B$ if
$N_G(B) \subseteq X$ and $X \subseteq B \cup N_G(B)$.
Note that a cap of $B$ is a potential root bag of a partial
tree-decomposition of $B$. 
For each block $B$, 
let $\calB_{\Pi}(B)$ denote the set of all caps of $B$ belonging
to $\Pi$. Recall that, for each vertex set $U \subseteq V(G)$, $\calC(U)$ denotes
the set of components of $G[U]$. The following recurrence holds.

\begin{eqnarray}
\textstyle \tw_{\Pi}(G, B) = \min_{X \in \calB_{\Pi}(B)} 
\max\{|X| - 1, \max_{C \in \calC(B \setminus X)} \tw_{\Pi}(G, C)\}\}
\label{eqn:recurrence}
\end{eqnarray}

BT dynamic programming evaluates this recurrence for blocks in the increasing order of cardinality and obtains $\tw_{\Pi}(G) = \tw_{\Pi}(G, V(G))$. Tracing back the
recurrences, we obtain a tree-decomposition $T \in \calT_{\Pi}(G)$ with
$w(T) = \tw_{\Pi}(G)$.

Tamaki's PID algorithm \cite{tamaki2019positive}, unlike the original algorithm
of Bouchitt\'{e} and Todinca \cite{bouchitte2001treewidth}, does not construct
$\Pi(G)$ before applying dynamic programming.
It rather uses the above recurrence to generate relevant blocks and PMCs.
More precisely, PID is for the decision problem whether $\tw(G) \leq k$ for
given $G$ and $k$ and it generates all blocks $C$ with $\tw(C, G) \leq k$
using the recurrence in a bottom up manner. We have $\tw(G) \leq k$ if and
only if $V(G)$ is among those generated blocks.

\paragraph*{Contractors and contractions}
To extend the notation $G / e$ of a contraction by an edge to a contraction by multiple edges,
we define contractors. A \emph{contractor} $\gamma$ of $G$ is a partition of $V(G)$ into
connected sets. For contractor $\gamma$ of $G$, the contraction of $G$ by $\gamma$,
denoted by $G / \gamma$, is the graph obtained from $G$ by contracting each
part of $\gamma$ to a single vertex, with the adjacency inherited from $G$.
For notational convenience, we also view a contractor $\gamma$ as a mapping from
$V(G)$ to $\{1, 2, \ldots, m\}$, the index set of the parts of the partition $\gamma$.
In this view, the vertex set of $G / \gamma$ is $\{1, 2, \ldots, m\}$ and
$\gamma(v)$ for each $v \in V(G)$ is the vertex of $G / \gamma$ into which 
$v$ is contracted. For each $w \in V(G / \gamma)$, $\gamma^{-1}(w)$ is the
part of the partition $\gamma$ that contracts to $w$. For $U \subseteq V(G / \gamma)$,
we define $\gamma^{-1}(U) = \bigcup_{w \in U} \gamma^{-1}(w)$.

\section{Main algorithm}
\label{sec:main}
The pseudo code in Algorithm~\ref{alg:main} shows the main iteration of 
our treewidth algorithm. It starts from a
greedy upper bound and repeatedly improves the upper bound by algorithm RTW.
The call $RTW(G, k, \Pi)$, where 
$\Pi \subseteq \Pi(G)$ and $\tw_\Pi(G) \leq k + 1$, decides
if $\tw(G) \leq k$.
If $\tw(G) \leq k$, it returns YES with certificate $\Pi' \subseteq \Pi(G)$ such that
$\tw_{\Pi'}(G) \leq k$; otherwise it returns NO with
certificate $H$, a minimal contraction of $G$ such that $\tw(H) = k + 1$.

\begin{algorithm}
\caption{Main iteration for computing $\tw(G)$}
\label{alg:main}
\begin{algorithmic}[1]
\Ensure compute $\tw(G)$ for given $G$
\State {$T \gets$ a minimal tree-decomposition of $G$ obtained by a greedy algorithm}
\State {$\Pi \gets $ the set of bags of $T$}
\State {$k \gets w(T)$}
\While {true}
  \State call $RTW(G, k - 1, \Pi)$
  \If {the call returns NO with certificate $H$}
     \State {stop: $\tw(G)$ equals $k$ with $\tw(G) \leq k$ certified by $\Pi$ and $\tw(G) \geq k$ certified by $H$}
  \Else
     \State{$k \gets k - 1$}
     \State{$\Pi \gets $ the certificate of the YES answer}
  \EndIf
\EndWhile
\end{algorithmic}
\end{algorithm}

The pseudo code in Algorithm~\ref{alg:basic} describes RTW in its
basic form. We sketch here the functions of subalgorithms used in this algorithm. 
More details can be found in subsequent sections.
\begin{algorithm}
\caption{Procedure $RTW(G, k, \Pi)$}
\label{alg:basic}
\begin{algorithmic}[1]
\Require 
$\Pi \subseteq \Pi(G)$ and 
$\tw_\Pi(G) \leq k + 1$
\Ensure returns YES with $\Pi \subseteq \Pi(G)$ such that $\tw_\Pi(G) \leq k$ if $\tw(G) \leq k$;
NO with a minimal contraction $H$ of $G$ such that $\tw(H) = k + 1$ otherwise
\State {create an HPID instance $s$ for $G$ and $k$}
\State {$s.\importPMCs(\Pi)$}
\If {$s.\width() \leq k$} 
  \State{ \return YES with $s.\usefulPMCs()$}
\EndIf
\State {order the edges of $G$ appropriately as $e_1$, $e_2$, \ldots $e_m$.}
\For{$i = 1, \ldots, m$}
  \State {$\Theta \gets \contractPMCs(s.\usefulPMCs(), G, e_i)$}
  \State call $RTW(G / e_i, k, \Theta)$
  \If {the call returns NO with certificate $H$} 
     \State {\return NO with certificate $H$}
  \Else 
    \State {$\Psi \gets$ the certificate for the YES answer}
    \State {$\Psi' \gets \uncPMCs(\Psi, G, e)$}
    \State {$s.\importPMCs(\Psi')$}
    \State {$s.\improve(UNIT\_BUDGET \times i)$}
    \If {$s.\width() \leq k$} 
      \State {\return YES with $s.\usefulPMCs()$}
    \EndIf
  \EndIf
\EndFor
\State $s.\finish()$
\If {$s.\width() \leq k$} 
\State{\return YES with $s.\usefulPMCs()$}
\Else 
\State {\return NO with certificate $G$}
\EndIf
\end{algorithmic}
\end{algorithm}

Our method of local search in the space of sets of PMCs is a heuristic variant,
which we call HPID, of the PID algorithm due to Tamaki \cite{tamaki2019positive}.
PID constructs partial tree-decompositions
of width $\leq k$ using the recurrence of BT dynamic programming in a bottom up
manner to exhaustively generate all partial tree-decompositions 
of width $\leq k$, so that we have a tree-decomposition of width $\leq k$
if and only if $\tw(G) \leq k$. HPID uses the same recurrence to 
generate partial tree-decompositions of width $\leq k$ but the
aim is to quickly generate a tree-decomposition of $G$ of width $\leq k$ and
the generation order it employs does not guarantee exhaustive generation.
The state of HPID computation is characterized by the set $\Pi$ of root bags
of the generated partial tree-decompositions. Recall that the bags of
the set of partial tree-decompositions generated by the BT recurrence are
PMCs, so $\Pi \subseteq \Pi(G)$. Using BT dynamic programming, we can
reconstruct the set of partial tree-decompositions from $\Pi$, if needed, in time
linear in $|\Pi|$ and polynomial in $|V(G)|$. Thus, we may view HPID
as performing a local search in the space of sets of PMCs. This view
facilitates communications between HPID and external upper bound heuristics.
Those communications are done through the following operations.

We consider each invocation of HPID as an entity having a state.
Let $s$ denote such an invocation instance of HPID for $G$ and $k$.
Let $\Pi(s)$ denote the set of PMCs that are root bag of the partial
tree-decompositions generated so far by $s$.
The following operations are available.
\begin{description}
    \item[$s.\width()$] returns $\tw_{\Pi(s)}(G)$.
    \item[$s.\usefulPMCs()$] returns the set of PMCs that are the root bags of
    the partial tree-decompositions of width $\leq s.width()$ generated so far by $s$.
    \item[$s.\importPMCs(\Pi)$] updates $\Pi(s)$ to $\Pi(s) \cup \Pi$ and updates the
    set of partial tree-decompositions by BT dynamic programming.
    \item[$s.\improve(\budget)$] generates more partial tree-decompositions under the specified budget, in terms of the number of search step spent for the generation.
    \item[$s.\finish()$] exhaustively generates remaining partial decompositions of
    width $\leq k$, thereby deciding if $\tw(G) \leq k$.
\end{description}
See Section~\ref{sec:hpid} for details of these procedures.

We use two additional procedures.
\begin{description}
    \item [$\uncPMCs(\Pi, G, e)$], where $e$ is an edge of $G$ and $\Pi \subseteq \Pi(G /e)$, 
    returns $\Pi' \subseteq \Pi(G)$ such that $\tw_{\Pi'}(G) \leq \tw_{\Pi}(G / e) + 1$ and
    possibly $\tw_{\Pi'}(G) \leq \tw_{\Pi}(G / e)$
    \item [$\contractPMCs(\Pi, G, e)$], where $e$ is an edge of $G$ and $\Pi \subseteq \Pi(G)$, 
    returns $\Pi' \subseteq \Pi(G / e)$ such that $\tw_{\Pi'}(G / e) \leq \tw_{\Pi}(G)$ and
    possibly $\tw_{\Pi'}(G / e) \leq \tw_{\Pi}(G) - 1$
\end{description}
See Sections~\ref{sec:uncontract} and \ref{sec:contract} for details of these procedures.

Given these procedures, RTW works as follows. It receives $G$, $k$, and $\Pi$
such that $\tw_{\Pi}(G) \leq k + 1$ and creates an HPID instance $s$ for $G$ and $k$
and let it import $\Pi$.
If it turns out that $\tw_{\Pi}(G) \leq k$ at this point, then
RTW returns YES with $s.\usefulPMCs()$, a subset $\Pi'$ of $\Pi$ such that
$\tw_{\Pi'}(G) \leq k$, as the certificate. Otherwise, it orders the edges of
$G$ in such a way to increase the chance of having an edge $e$ width $\tw(G / e) = k + 1$
early in the list if any. Then it iterates over those edges.
To process $e_i$ it makes a recursive call $RTW(G / e_i, k, \Theta)$
where $\Theta \subseteq \Pi(G / e)$ is obtained by "contracting" $\Pi$.
If the result is negative, the answer of $RTW(G, k , \Pi)$ is also negative with
the same certificate. If the result is positive with $\Psi \subseteq \Pi(G / e_i)$,
then $\Psi$ is "uncontracted" to $\Psi' \subseteq \Pi(G)$, which is imported to $s$.
Then it lets $s$ advance its PID state under a budget proportional to $i$. If $s$ succeeds
in finding tree-decompositions of $G$ of width $k$, then RTW returns YES
with the certificate constructed by $s$. Otherwise, it proceeds to the next edge.
When it has tried all edges without resolving the question, it lets $s$ finish
the exhaustive generation of partial tree-decompositions to answer the question.
If it turns out that $\tw(G) \leq k$, it returns YES with the certificate provided by
$s$. Otherwise it returns NO with the certificate being $G$ itself.

The correctness of this algorithm can be proved by straightforward induction and
does not depend on the procedures $\expand$, $\contractPMCs$, or $\uncPMCs$ except
that the procedure $\contractPMCs(\Pi, G, e)$ must return $\Theta$ such that 
$\tw_{\Theta}(G /e) \leq \tw_{\Pi}(G)$ as promised. 
On the other hand, practical efficiency of this algorithm heavily depends on
the performances of these procedures. If they collectively work really well,
then we expect that the \textbf{for} loop would exit after trying only a
few edges, assuming $\tw(G) \leq k$, and $s.\finish()$ would be called only if
$\tw(G) = k + 1$ and $\tw(G / e) \leq k$ for every edge $e$. On the other extreme
of perfect incapability of these procedures, 
the \textbf{for} loop would always run to the end and $s.\finish()$ would be
called in every call of $RTW(G, k, \Pi)$, making the recursion totally meaningless.
Our efforts are devoted to developing effective methods for these procedures.

\section{Heuristic PID}
\label{sec:hpid}
In this section, we give some details of the HPID algorithm.
In particular, we describe in some details how the procedures 
$\improve(\budget)$ and $\finish()$ work.

We first describe how we use Recurrence~\ref{eqn:recurrence} to generate a
new partial tree-decomposition from existing ones. 
The method basically follows that of PID \cite{tamaki2019positive} but
there are some differences. The most important difference is in the manners
we turn tree-decompositions into rooted tree-decompositions, which is done
in order to restrict partial tree-decompositions to be generated. 
In the original PID, the choice of roots heavily depends on the total order assumed on
$V(G)$. For the sake of interactions of HPID with other upper bound components
through PMCs, we prefer the choice to depend less on the vertex order and
thus be fairer for vertices.

Fix $G$ and $k$. We assume a total order $<$ on $V(G)$ and 
say that $U \subseteq V(G)$ is \emph{larger} then $V \subseteq V(G)$ if
$|U| > |V|$ or $|U| = |V|$ and $U$ is lexicographically larger than $V$.
We say that a block $B$ of $G$ is \emph{feasible} if
$\tw(B, G) \leq k$. We use recurrence~\ref{eqn:recurrence}, with $\Pi$
set to $\Pi(G)$, to generate feasible blocks. Our goal is to see if
$V(G)$ is feasible and, to this end, it turns out that we do not need
to generate all feasible blocks: it suffices to generate only small 
feasible blocks except for $V(G)$ itself, where a block $B$ is \emph{small} if
there is some block $B'$ with $N_G(B') = N_G(B)$ such that $B' > B$.

To see this, we construct rooted tree-decompositions from
minimal triangulations of $G$.
Let $H$ be a minimal triangulation of $G$. We define a rooted tree
$D_H$ on the set of maximal cliques of $H$, which may be denoted by 
$\Pi(H)$ because every PMC of $H$ is a maximal clique.
For $X \in \Pi(H)$ and a minimal separator $S \subset X$, let $B(S, X)$
denote the full component of $S$ that intersects $X$. Note that 
$B(S, X)$ is a block since $N_G(B(S, X)) = S$ is a minimal separator.
For $X, Y \in \Pi(H)$ such that $X \cap Y$ is a separator of $H$,
let $S(X, Y)$ denote the inclusion-minimal minimal separator of
$H$ contained in $X \cap Y$. Such an inclusion minimal separator is unique:
if distinct $S_1, S_2 \subseteq X \cap Y$ are both inclusion-minimal separators,
then both of the strict inclusions of $B(S_1, X) \subset B(S_2, Y)$ and
$B(S_2, Y) \subset B(S_1, X)$ must hold, which is impossible.

We first define a dag $W_H$ on $\Pi(H)$: for distinct $X, Y \in \Pi(H)$there, 
$W_H$ has an edge from $X$ to $Y$ if $X \cap Y$ is a separator of $H$
and $B(S(X, Y), Y)$ is larger than $B(S(X, Y), X)$. 
\begin{proposition}
    $W_H$ is acyclic.
\end{proposition}
\begin{proof}
  Suppose, for contradiction, there is a directed cycle in $W_H$ and
  let $X_1$, \ldots, $X_m$, $X_{m+1} = X_1$ be the shortest such.
  Let $S = S(X_1, X_2)$.
  It cannot be that $m = 2$, since then we would have both
  $B(S, X_1) < B(S, X_2)$ and $B(S, X_1) > B(S, X_2)$.

  Let $i \geq 2$ be such that
  $X_i \not\subseteq B(S, X_1) \cup S$ and $X_{i + 1} \subseteq B(S, X_1) \cup S$.
  Such $i$ must exist since $X_2 \not\subseteq B(S, X_1) \cup S$ and 
  $X_{m + 1} \subseteq B(S, X_1) \cup S$. Let $S' = S(X_i, X_{i + 1})$.
  Since every block of $H$ is either contained in $B(S, X_1)$ or disjoint from it,
  we have $B(S', X_i) \cap B(S, X_1) = \emptyset$ and
  $B(S', X_{i + 1}) \subseteq B(S, X_1)$. Since $S'$ separates these blocks,
  we must have $S' \subseteq N_G(B(S, X_1)) = X_1 \cap X_2$. 
  Since $S$ and $S'$ are both inclusion-minimal, we must have
  $S = S'$ as argued above. Then, we have 
  $B(S, X_2) > B(S, X_1) \supseteq B(S, X_{i + 1}) > B(S, X_i)$ and 
  therefore we have an edge from $X_i$ to $X_2$, contradicting
  the assumption that our directed cycle is the shortest.
\end{proof}

Now we construct a directed tree $D_H$ on $\Pi(H)$ with a unique sink.
As $W_H$ is acyclic, it has a sink $X_0$.
Let $\calB$ denote the set of components of $G \setminus X_0$.
Each $B \in \calB$ is a block since $N_H(B) \subseteq X_0$ is a minimal-separator.
Let $\Pi(H, B)$ denote the set of maximal cliques of $H$ contained
in $B \cup N_H(B)$. Note that $\Pi(H, B)$, $B \in \calB$, partitions
$\Pi(H) \setminus \{X_0\}$. For each such $B$, we construct
a directed tree $D_H(B)$ on $\Pi(H, B)$ with unique sink $X_B$
such that $W_H$ has an edge from $X_B$ to $X_0$. 
Combining $D_H(B)$, $B \in \calB$, with these edges from $X_B$ to $X_0$,
we obtain $D_H$. It remains to show how we construct $D_H(B)$.

Observe that every $B \in \calB$ is small.
For each small block $B$, we construct a directed tree $D_H(B)$
on $\Pi(H, B)$ with sink $X_B$ such that $N_H(B) \subseteq X_B$ inductively
as follows.  Let $\calC_B$ denote the set of caps of $B$ belonging to 
$\Pi(H)$. By the definition of caps, each $X \in \calC_B$ satisfies $N_H(B) \subseteq X$
and $X \subseteq B \cup N_H(B)$. The subgraph of $W_H$ induced by $\calC_B$
has a sink $X_B$ since $W_H$ is acyclic. Let $\calB(X_B, B)$ denote the
set of blocks of $H$ that are components of $B \setminus X_B$.
For each $B' \in \calB(X_B, B)$, we have $B' \subseteq B$ and, moreover, 
for each block $C \neq B$ of $N_H(B)$, we have $C \subseteq C(N_H(B'), X_B)$.
Therefore, since there is a block $C$ of $N_H(B)$ such that $C > B$ as
$B$ is small, we have $C(N_H(B'), X_B) > B'$ for each $B'$, that is, $B'$ is small.
By the induction hypothesis, we have a directed tree $D_H(B')$ on $\Pi(H, B')$
with sink $X_{B'}$ such that $N_H(B') \subseteq X_{B'}$, for each $B' \in \calB(X_B, B)$.
Combining $D_H(B')$, $B' \in \calB(X, B)$, with an edge from each $X_{B'}$ to $X_B$,
we obtain the desired directed tree $D_H(B)$.

Let $H$ be a minimal triangulation of $G$ such that $\tw(H) = \tw(G)$.
In view of the existence of the rooted clique tree $D_H$ of $H$,
feasibility of $V(G)$ can be determined by generating only small feasible blocks
using recurrence~\ref{eqn:recurrence} and then seeing if the same recurrence
can be used to show $\tw(G) = \tw(V(G), G) = k$. Thus, each 
HPID instance $s$ maintains a set set $\calF$ of small feasible blocks. 
To generate a new feasible block to add to $\calF$,
it invokes a backtrack search procedure $\searchNewFeasible(B)$ 
on a block $B \in \calF$ which enumerates 
$\calB \subseteq \calF$ such that 
\begin{enumerate}
\item $B \in \calB$ and $B$ is the largest block in $\calB$ and
\item there is a block $B_\calB$ that is either small or is equal to 
$V(G)$ and a PMC $X_\calB \in \Pi(G)$ such that
$\calC(B_\calB \setminus X_\calB) = \calB$.
\end{enumerate}
For each such $\calB$ found, we add $B_\calB$ to $\calF$ since
the recurrence~\ref{eqn:recurrence} shows that $B_\calB$ is feasible.

Procedure $s.\improve(\budget)$ uses this search procedure as follows.
It uses a priority queue $Q$ of small feasible blocks, in which 
larger blocks are given higher priority. It first put all blocks 
in $\calF$ to $Q$. Then, it dequeues a block $B$,
call $\searchNewFeasible(B)$, and add newly generated feasible blocks to $Q$.
This is repeated until either $Q$ is empty or the cumulative number of
search steps exceeds $\budget$. Because of the queuing policy, there is
a possibility of $V(G)$ found feasible, when it is indeed feasible, 
even with a small budget.

Procedure $s.\finish()$ works similarly, except that smaller blocks 
are given higher priority in the queue and the budget is unlimited,
to generate all small feasible blocks and $V(G)$ if it is feasible.

An alternative way to to implement the $\finish()$ procedure is to
call another exact treewidth algorithm based on BT dynamic programming,
such as SemiPID \cite{tamaki2019computing}, to decide if $V(G)$ is feasible.
The implementation used in our experiment uses this alternative method.

\section{Minimalizing tree-decompositions}
\label{sec:minimalize}
Given a graph $G$ and a triangulation $H$ of $G$, \emph{minimalizing} $H$
means finding a minimal triangulation $H'$ of $G$ such that $E(H') \subseteq E(H)$.
Minimalizing a tree-decomposition $T$ of $G$ means finding
a minimal tree-decomposition $T'$ of $G$ whose bags are maximal cliques of
the minimalization of $\fillg(G, T)$.
We want to minimalize a tree-decomposition for two reasons. One is our decision to represent a set of tree-decompositions 
by a set of PMCs. Whenever we get a tree-decomposition $T$ by some method that may produce non-minimal tree-decompositions,
we minimalize it to make all bags PMCs. Another reason is that minimalization may reduce the width.
We have two procedures for minimalization. When the second reason is of no concern, we use $\minimalize(T)$ which is an implementation of one of the standard triangulation minimalization algorithm due to Blair {\it et al} \cite{blair2001practical}.
When the second reason is important, we use $\minopt(T)$, which finds a minimalization of $T$ of the smallest width. This task is NP-hard, but the following algorithm works well in practice.

Say a minimal separator of $G$ is \emph{admissible for} $T$ if it is a clique of $\fillg(G, T)$. Observe that, for every minimalization $T'$ of $T$, every separator used by $T'$ is a minimal separator of $G$ admissible for $T$.
We first construct the set of all minimal separators of $G$ admissible for $T$. Then we apply the SemiPID variant of BT dynamic programming, due to Tamaki \cite{tamaki2019computing}, to this set and obtain a tree-decomposition of the smallest width, among those using only admissible minimal separators. Because of the admissibility constraint, the number of minimal separators is much smaller  and both the enumeration part and the SemiPID part run much faster in practices than in the general case without such constraints.

\section{Uncontracting PMCs}
\label{sec:uncontract}
In this section, we develop an algorithm for procedure $\uncPMCs(G, \Pi, e)$.
In fact, we generalize this procedure to $\uncPMCs(G, \Pi, \gamma)$, 
where the third argument is a general contractor of $G$.

Given a graph $G$, $\Pi \subseteq \Pi(G)$, and a contractor $\gamma$ of $G$,
we first find tree-decompositions $T \in \calT_{\Pi}$ that minimize 
$w(\gamma^{-1}(T))$. This is done by BT dynamic programming
over $\calT_\Pi(G / \gamma)$, using bag weights defined as follows.
For each weight function $\omega$ that assigns weight $\omega(U)$ to
each vertex set $U$, define the width of tree-decomposition $T$
with respect to $\omega$, denoted by $\tw(G, \omega)$, to be 
the maximum of $\omega(X)$ over all bags of $T$.
Thus, if $\omega$ is defined by $\omega(U) = |U| - 1$ then $\tw(G, \omega) = \tw(G)$.
A natural choice for our purposes is to set $\omega(X) = |\gamma^{-1}(X)| - 1$. 
Then, the width
of a tree decomposition $T$ of $G / \gamma$ with respect to this bag weight
is $w(\gamma^{-1}(T))$. Therefore, BT dynamic programming with this weight function
$\omega$ gives us the desired tree-decomposition in $\calT_\Pi(G / r)$.

We actually use a slightly modified weight function, considering the possibility
of reducing the weight of $\gamma^{-1}(T)$ by minimalization.

Let $T \in \calT_{\Pi}(G / \gamma)$ and $X$ a bag of $T$. If
$X' = \gamma^{-1}(X)$ is a PMC of $G$, then every minimalization 
of $\gamma^{-1}(T)$ must contain $X'$ as a bag. Therefore, 
if $|X'| > k + 1$ then it is impossible that
the width of $\gamma^{-1}(T)$ is reduced to $k$ by minimalization.
On the other hand, if $X'$ is not a PMC, then no minimalization of
$\gamma^{-1}(T)$ has $X'$ has a bag and there is a possibility
that there is a minimalization of $\gamma^{-1}(T)$ of width $k$ even
if $|X'| > k + 1$. These considerations lead to the following definition
of our weight function $\omega$.
\begin{eqnarray}
\omega(U) & = & 2|\gamma^{-1}(U)| \hspace{1in} \mbox{if $\gamma^{-1}(U)$ is a PMC of $G$} 
\label{eqn:1}\\
\omega(U) & = & 2|\gamma^{-1}(U)| - 1 \hspace{1in}\mbox{otherwise} \label{eqn:2}
\end{eqnarray}

Algorithm~\ref{alg:uncontract} describes the main steps of procedure $\uncPMCs(\Pi, G, \gamma)$.

\begin{algorithm}
\caption{Procedure $\uncPMCs(\Pi, G, \gamma)$}
\label{alg:uncontract}
\begin{algorithmic}[1]
\Require 
$\Pi \subseteq \Pi(G / \gamma)$
\Ensure returns $\Pi' \subseteq \Pi(G)$ that results from uncontracting $\Pi$ and
then minimalizing
\State let $\omega$ be the weight function on $2^{V(G / \gamma)}$ defined by
equations~\ref{eqn:1} and ~\ref{eqn:2}
\State use BT dynamic programming to obtain tree-decompositions $T_i$, $1 \leq i \leq m$, of $G / \gamma$ such that $w(T_i, \omega) = \tw_{\Pi}(G, \omega)$ 
\For {each $i$, $1 \leq i \leq m$} 
  \State $T'_i \gets \minopt(\gamma^{-1}(T_i))$
  \State $\Pi_i \gets$ the set of bags of $T'_i$
\EndFor
\State \return $\bigcup_i \Pi_i$
\end{algorithmic}
\end{algorithm}

\section{Contracting PMCs}
\label{sec:contract}
The algorithm for procedure $\contractPMCs$ is similar to that for 
$\uncPMCs$. Given a graph $G$, $\Pi \subseteq \Pi(G)$, and a contractor $\gamma$ of $G$,
we first find tree-decompositions $T \in \calT_{\Pi}(G / \gamma)$ that minimize 
$w(\gamma(T))$. This is done by BT dynamic programming with the following weight
function $\omega$.
\begin{eqnarray*}
\omega(U) & = & 2|\gamma(U)| \hspace{1in} \mbox{if $\gamma(U)$ is a PMC of $G / \gamma$} \\
\omega(U) & = & 2|\gamma(U)| - 1 \hspace{1in}\mbox{otherwise}
\end{eqnarray*}
Then, we minimalize those tree-decompositions and collect the bags of those
minimalized tree-decompositions.

\section{Safe separators}
\label{sec:safe_sep}

Bodlaender and Koster \cite{bodlaender2006safe} introduced
the notion of safe separators for treewidth.
Let $S$ be a separator of a graph $G$.
We say that $S$ is \emph{safe for treewidth}, or simply safe, 
if $\tw(G) = \tw(G \cup K(S))$.
As every tree-decomposition of $G \cup K(S)$ must have a bag containing $S$,
$\tw(G)$ is the larger of $|S| - 1$ and $\max \{\tw(G[C \cup N_G(C)] \cup K(N_G(C))\}$,
where $C$ ranges over all the components of $G \setminus S$. Thus, the task of
computing $\tw(G)$ reduces to the task of computing $\tw(G[C \cup N_G(C)] \cup K(N_G(C))\}$ for every component $C$ of $G \setminus S$.
The motivation for looking at safe separators of a graph is that
there are sufficient conditions for a separator being safe and
those sufficient conditions lead to an effective preprocessing method for
treewidth computation. We use the following two  sufficient conditions.

A vertex set $S$ of $G$ is an \emph{almost-clique} if $S \setminus \{v\}$
is a clique for some $v \in S$.
Let $R$ be a vertex set of $G$. A contractor $\gamma$ of $G$ is \emph{rooted on $R$}
if, for each part $C$ of $\gamma$, $|C \cap R| = 1$.

\begin{theorem}
\label{thm:safe}
Bodlaender and Koster \cite{bodlaender2006safe} 
\begin{enumerate}
\item If $S$ is an almost-clique minimal separator of $G$, then $S$ is
safe.
\item Let $lb$ be a lower bound on $\tw(G)$. Let $C \subseteq V(G)$ be connected
and let $S = N_G(C)$. Suppose
(1) $\tw(G[C \cup S] \cup K(S)) \leq lb$ and
(2) $G[C \cup S]$ has a contractor $\gamma$ rooted on $S$ such that
$G[C \cup S] / \gamma$ is a complete graph.
Then, $S$ is safe.
\end{enumerate}
\end{theorem}

We use safe separators both for preprocessing and during recursion.
For preprocessing, we follow the approach of \cite{tamaki2021heuristic}:
to preprocess $G$, we fix a minimal triangulation $H$ of $G$ and
test the sufficient conditions in the theorem for each minimal separator of $H$.
Since deciding if the second condition holds is NP-complete, we use a heuristic
procedure. 
Let $\calS$ be the set of all minimal separators of $H$ that are confirmed to
satisfy the first or the second condition of the theorem. Let $\calA$ be a tree-decomposition of $G$ that uses all separators of $\calS$ but no other separators. Then, $\calA$ is what is called a \emph{safe-separator decomposition} in \cite{bodlaender2006safe}. 
A tree-decomposition of $G$ of width $\tw(G)$ can be obtained from $\calA$ by
replacing each bag $X$ of $\calA$ by a tree-decomposition of
$G[X] \cup \bigcup_{C \in \calC(G \setminus X)} K(N_G(C))$,
the graph obtained from the subgraph of $G$ induced by $X$
by filling the neighborhood of every component of
$G \setminus X$ into a clique.

Safe separators are also useful during the recursive computation.
Given $G$, we wish to find a contractor $\gamma$ of $G$ such that $\tw(G / \gamma) = \tw(G)$,
so that we can safely recurse on $G / \gamma$. The second sufficient condition in 
Theorem~\ref{thm:safe} is useful for this purpose. Let $C$, $S$, and $\gamma$ be as
in the condition. We construct $\gamma'$ such that $\tw(G / \gamma') = \tw(G)$ as follows.
The proof of this sufficient condition is based on the fact that we get a clique on
$S$ when we apply the contractor $\gamma$ on $G[C \cup S]$. 
Thus, we may define a contractor $\gamma'$ on $G$ such that
$G / \gamma' = (G \setminus C) \cup K(S)$. As each tree-decomposition of
$\tw(G / \gamma)$ can be extended to a tree-decomposition of $G$,
using the tree-decomposition of $G[C \cup S] \cup K(S)$ of width at most $lb \leq \tw(G)$,
we have $\tw(G / \gamma')$ = $\tw(G)$ as desired.
When the recursive call on $\tw(G / \gamma')$ returns a certificate
$\Pi \subseteq \Pi(G / \gamma')$ such that $\tw_\Pi(G / \gamma') \leq k$,
we need to "uncontract" $\Pi$ into a $\Pi' \subseteq \Pi(G)$
such that $\tw_{\Pi'}(G) \leq k$. Fortunately, this can be done without
invoking the general uncontraction procedure. Observe first that each PMC
in $\Pi$ naturally corresponds to a PMC of $(G \setminus C) \cup K(S)$, which in 
turn corresponds to a PMC of $G$ contained in $V(G) \setminus C$.
Let $\Pi_1$ be the set of those PMCs of $G$ to which a PMC in $\Pi$ corresponds in that manner.
Let $\Pi_2 \subseteq \Pi(G[C \cup S] \cup K(S))$ be such that 
$\tw_{\Pi_2}(G[C \cup S] \cup K(S)) \leq lb$. Similarly as above, 
each PMC of $\Pi_2$ corresponds to a PMC of $G$ contained $C \cup S$.
Let $\Pi'_2$ denote the set of those PMCs of $G$ to which a PMC in $\Pi_2$ corresponds.
As argued above, a tree-decomposition in $\calT_{\Pi}((G \setminus C) \cup K(S))$ of
$(G \setminus C) \cup K(S)$ and a tree-decomposition in $\calT_{\Pi_2}(G[C \cup S] \cup K(S))$ of
$G[C \cup S] \cup K(S)$ can be combined into a tree-decomposition belonging to
$\calT_{\Pi'_2}(G)$ of width $\leq k$. Thus, $\Pi'_2$ is a desired certificate for
$\tw(G) \leq k$.

\section{Edge ordering}
\label{sec:edge_order}
We want an
edge $e$ such that $\tw(G / e) = \tw(G)$,
if any, to appear early in our edge order. 
Heuristic criteria for such an ordering have been studied in the classic work on contraction based lower bounds \cite{bodlaender2011treewidth}.
Our criterion is similar to those but differs in that it derives from a special case 
of safe separators. The following is simple corollary of Theorem~\ref{thm:safe}.

\begin{proposition}
\label{prop:almost_clique_contraction}
Let $e = \{u, v\}$ be an edge of $G$ and let $S = N_G(v)$.
Suppose $S \setminus \{u\}$ is a clique of $G$.
Then, we have $\tw(G / e) = \tw(G)$.
\end{proposition}
If $e$ satisfies the above condition, then we certainly put $e$ 
first in the order. 
Otherwise, we evaluate  $e$ in terms of its closeness
to this ideal situation. 
Define the \emph{deficiency} of graph $H$, denoted by
$\defic(H)$, to be the number of edges of its complement graph.
For each ordered pair $(u, v)$ of adjacent vertices of $G$,
let $\defic_G(u, v)$ denote 
$\defic(G[N_G(v) \cup \{v\}] / \{u, v\})$.
Note that $\defic_G(u, v) = 0$ means 
that the condition of the above
proposition is satisfied with $S = N_G(v)$. Thus, we regard
$e = \{u, v\}$ preferable if either $\defic_G(u, v)$ or $\defic_G(v, u)$ is small.
We relativize the smallness with respect to 
the neighborhood size,
so the \emph{value} of edge $e = \{u, v\}$ is
$\min\{defic_G(u, v) / |N_G(v)|, defic_G(v, u) / |N_G(u)|\}$.
We order edges so that this value is non-decreasing.

\section{Suppressed edges}
\label{sec:suppressed}
Consider the recursive call on $G / e$ from the call of RTW on $G$,
where $e$ is an edge of $G$. Suppose there is an ancestor
call on $G'$ such that $G = G' / \gamma$ and edge $e'$ of $G'$
such that $\gamma$ maps the ends of $e'$ to the ends of $e$.
If the call on $G' / e'$ has been made and it is known that
$\tw(G' / e') \leq k$ then we know that $\tw(G / e) \leq k$,
since $G / e$ is a contraction of $G' / e'$.
In this situation, we that $e$ is \emph{suppressed} by the pair $(G', e')$.
We may omit the recursive call on $G / e$ without compromising the
correctness if $e$ is suppressed. For efficiency, however, it is preferable
to obtain the certificate $\Pi \subseteq \Pi(G / e)$ for
$\tw(G/ e) \leq k$
and feed the uncontraction of $\Pi$ to the HPID instance on $G$ to
help progress. Fortunately,
this can be done without making the recursive call on $G$ as follows.
Suppose $e$ is suppressed by $(G', e')$ and let $\Pi' \subseteq \Pi(G' / e')$
such that $\tw_{\Pi'}(G' / e') \leq k$. Let $\gamma'$ be 
the contractor of $G' / e'$ such that $G' / e'/ \gamma' = G / \gamma / e$:
such $\gamma'$ is straightforward to obtain from $\gamma$.
Letting $\Pi = \contractPMCs(\Pi, G' / e', \gamma')$, we obtain
$\Pi \subseteq \Pi(G / e)$ such that
$\tw_{\Pi}(G/ e) \leq k$.

\section{Experiments}
\label{sec:experiments}
We have implemented RTW and evaluated it by experiments.
The computing environment for our experiments is as follows.
CPU: Intel Core i7-8700K, 3.70GHz; RAM: 64GB; 
Operating system: Windows 10Pro, 64bit; 
Programming language: Java 1.8; JVM: jre1.8.0\_271.
The maximum heap size is set to 60GB. The implementation uses a
single thread except for additional
threads that may be invoked for garbage collection by JVM. 

Our primary benchmark is the bonus instance set of the exact treewidth track of PACE 2017 algorithm implementation challenge \cite{dell2018pace}. 
This set, consisting of 100 instances, 
is intended to be a challenge for future implementations
and, as a set, are hard for the winning solvers of the competition.
Using the platform of the competition, 
about half of the instances took more than one hour to solve 
and 15 instances took more than a day or were not solvable at all.

\begin{figure}[htbp]
\begin{center}
\includegraphics[width=5in]{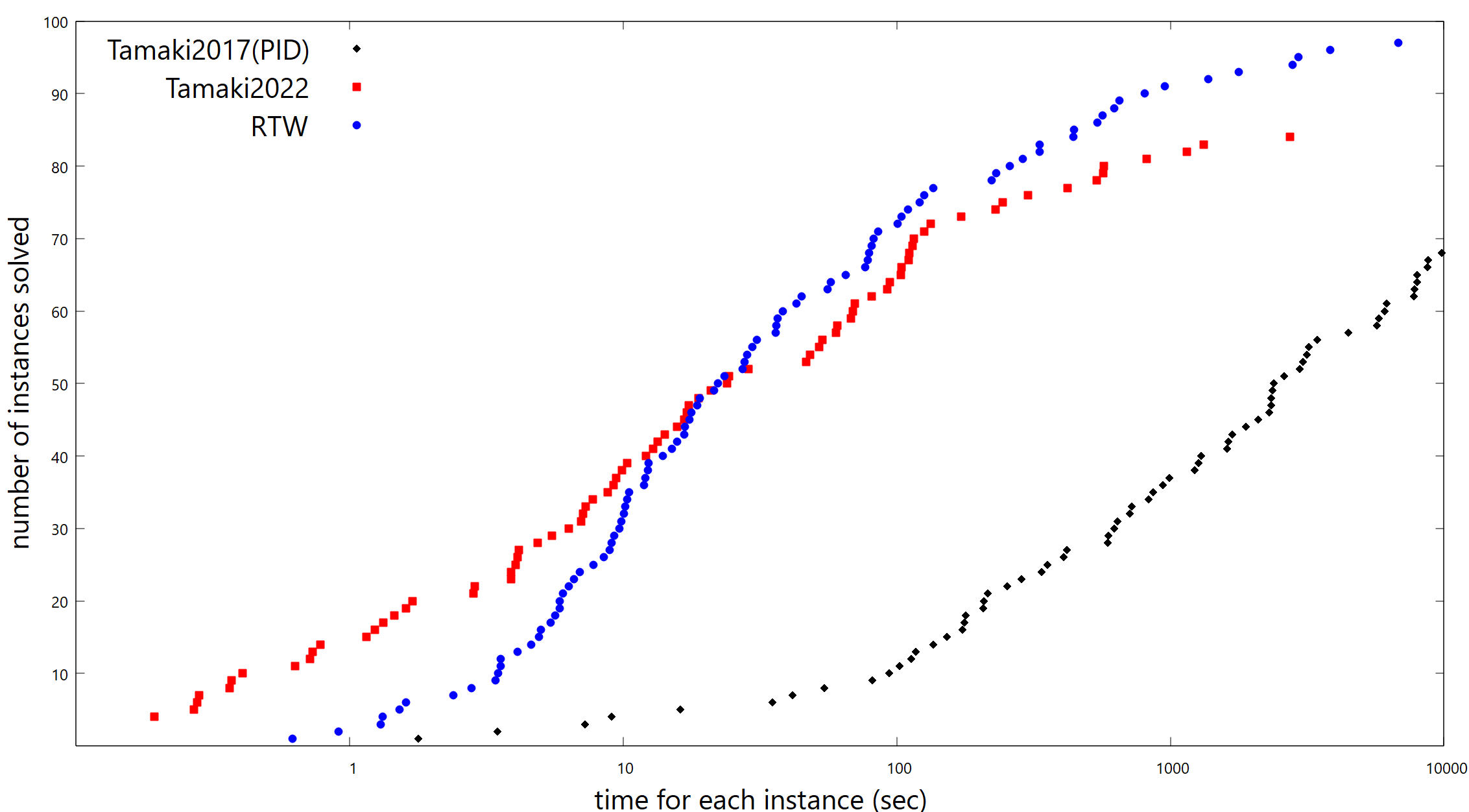}
\end{center}
\caption{Number of bonus instances solved within a specified time}
\label{fig:bonus_time_count}
\end{figure}

We have run our implementation on these instances with the timeout of
10000 seconds each. For comparison, we have run Tamaki's PID solver \cite{tamaki2019positive}, which is 
one of the PACE 2017 winners,
available at \cite{pid2017} and his new solver \cite{tamaki:LIPIcs.SEA.2022.17}
available at \cite{twalgorHEX}.
Figure~\ref{fig:bonus_time_count} summarizes the results on the bonus set.
In contrast to PID solver which solves only 68 instances within the timeout, RTW solves 98 instances.
Moreover, it solve 72 of them in 100 seconds and 92 of them in 1000 seconds. 
Thus, we can say that our algorithm drastically extends the scope of practically solvable instances.
Tamaki's new solver also quickly solves many instances that are hard for PID solver
and is indeed faster then RTW on many instances. However, its performance in terms of the number
of instances solvable in practical time is inferior to RTW.

\begin{figure}[htbp]
\begin{center}
\includegraphics[width=5in]{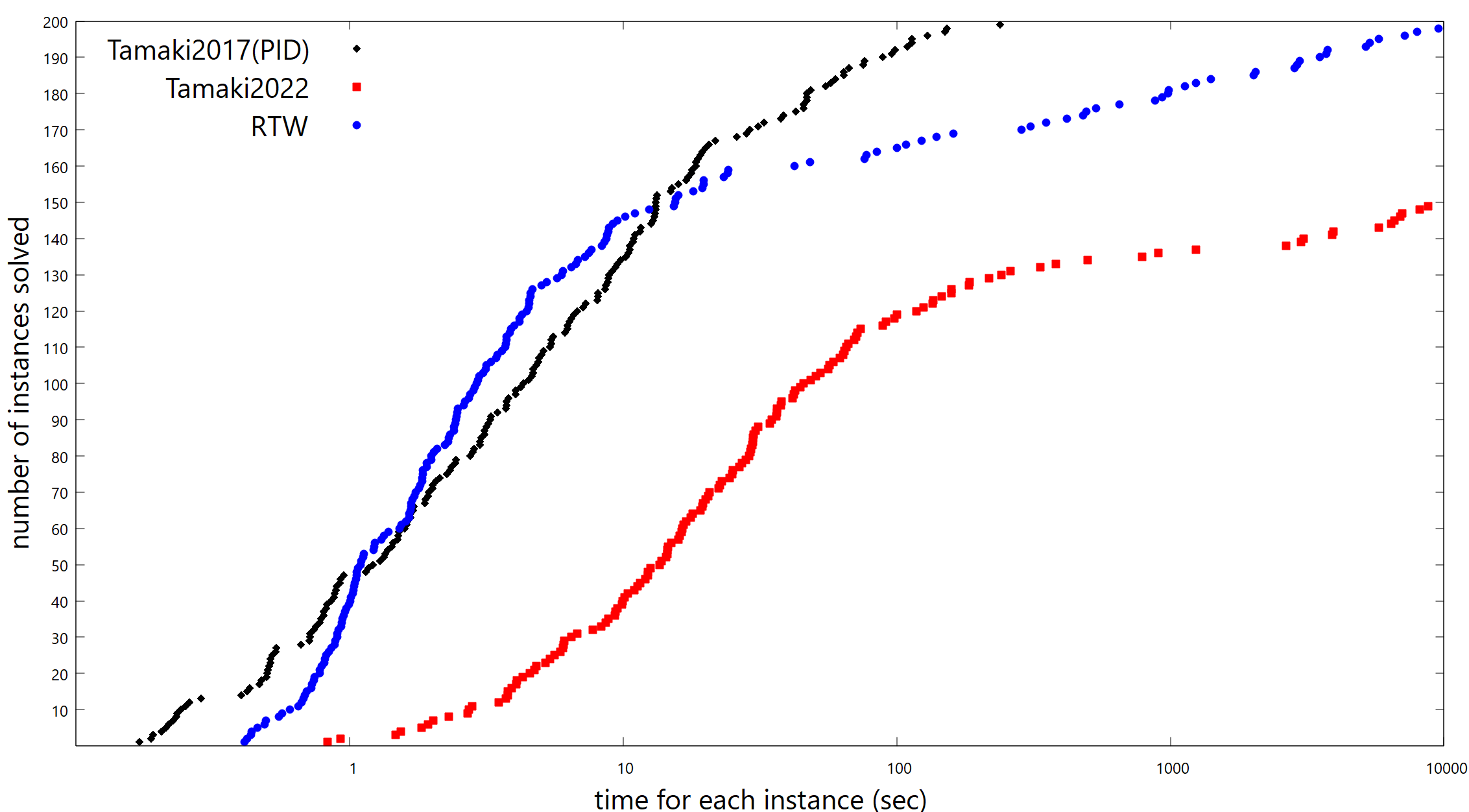}
\end{center}
\caption{Number of competition instances solved within a specified time}
\label{fig:pace17_time_count}
\end{figure}

We have also run the solvers on the competition set of the exact treewidth 
track of PACE 2017.
This set, consisting of 200 instances, is relatively easy and the two winning
solvers of the competitions solved all of the instances within the allocated timeout
of 30 minutes for each instance. 
Figure~\ref{fig:pace17_time_count} summarizes the results on the competition set.
Somewhat expectedly, PID performs the best on this instance set. 
It solves almost all instances in 200 seconds for each instance, while RTW fails to do so on about 30 instances.
There are two instances that RTW fails to solve in 10000 seconds and one instance
it fails to solve at all. Tamaki's new solver shows more weakness on this set, failing to
sove about 50 instances in the timeout of 10000 seconds.

These results seem to suggest that RTW and PID should probably complement
each other in a practical treewidth solver. 

\section{Conclusions and future work}
\label{sec:conclusions}
We developed a treewdith algorithm RTW that works recursively on contractions.
Experiments show that our implementation solves many instances in practical time that are hard to solve for
previously published solvers. RTW, however, does not perform well on some instances
that are easy for conventional solvers such as PID. A quick compromise would be to
run PID first with an affordable timeout and use RTW only when it fails. 
It would be, however, interesting and potentially fruitful to closely examine those instances 
that are easy for PID and hard for RTW and, based on such observations, to look for
a unified algorithm that avoids the present weakness of RTW.

\bibliography{main}
\end{document}